\documentclass[letterpaper,english,aps,prl,twocolumn,floatfix, showpacs, amsfonts, amssymb]{revtex4-1}
\usepackage{epsfig, graphicx,psfrag,amsmath,amssymb,float}
\usepackage[T1]{fontenc}
\usepackage[latin9]{inputenc}
\usepackage{amsmath}
\usepackage{amssymb}

\usepackage{amscd}
\usepackage{bm}
\usepackage{psfrag}
\usepackage{appendix}
\usepackage{babel}

\newcommand{\be}{\begin{equation}}
\newcommand{\ee}{\end{equation}}
\newcommand{\bea}{\begin{eqnarray}}
\newcommand{\eea}{\end{eqnarray}}

\newcommand{\mc}{\mathcal}
\newcommand{\mb}{\mathbf}

\begin{document}

\title{Tkachenko polarons in vortex lattices}
\author{M. A. Caracanhas}
\author{V. S. Bagnato}
\author{R. G. Pereira}
\affiliation{Instituto de F\'{i}sica de S\~ao Carlos, Universidade de S\~ao Paulo, C.P. 369, S\~ao Carlos, SP,  13560-970, Brazil}

\date{\today}
\begin{abstract}

We  analyze the properties of impurities immersed in a  vortex lattice formed by ultracold bosons in the mean field quantum Hall regime. In addition to the effects of a  periodic lattice potential, the impurity is dressed by collective modes with parabolic dispersion (Tkachenko modes). We derive the effective polaron model, which contains a marginal impurity-phonon interaction. The polaron spectral function exhibits a Lorentzian broadening for arbitrarily small wave vectors even at zero temperature, in   contrast with the result for optical or acoustic phonons.  The unusually strong damping of Tkachenko polarons could  be detected experimentally using momentum-resolved spectroscopy. 

 \end{abstract}


\maketitle

The problem of a single particle propagating in a polarizable medium has been a subject of interest since the early days of  condensed matter physics and quantum field theory \cite{feynman}. The original motivation to study  the resulting quasiparticle, known as a polaron,  was to investigate the effects of the electron-phonon interaction in transport properties of solids \cite{devreese}. However, the concept is readily generalized if one replaces   electrons by mobile impurities and lattice phonons by collective modes of a dynamical background. Following this idea, ultracold atoms  have emerged as the new testing ground for polaron physics, with    proposed experimental setups  involving homogeneous Bose-Einstein condensates (BECs) \cite{bruderer}, dipolar molecules in self-assembled and optical lattices \cite{pupillo,herrera} and imbalanced Fermi gases \cite{schirotzek}. In the  fermionic case, a particularly exciting result is the recent observation of long-lived repulsive polarons \cite{koschorrek}.
 
Cold atom realizations of polarons may give insight  into long-standing problems in many-body phenomena, such as itinerant ferromagnetism \cite{koschorrek}. Furthermore, they  raise questions about what types of particle-boson interactions are possible and whether these interactions may lead to deviations from standard polaronic behaviour. For instance, it has been shown that a sharp transition in ground state properties can occur in  models  where the coupling depends on the particle momentum \cite{marchand, herrera}. Besides more general couplings, one may wonder how polaron properties get modified if  the  dispersion relation $\omega_{\bf q}$ of the collective mode differs from that of      conventional  optical or acoustic phonons (in the former case,  $\omega_{\bf q} =$ const., as  in the  Holstein model \cite{devreese}; in the latter, $\omega_{\bf q }=v_s  q$ for $ q\to0$, where $v_s$ is the sound velocity). 

It is known that vortex lattices in rapidly rotating two-dimensional BECs \cite{fetter,cooper} host collective modes with   quadratic dispersion $\omega_{\bf q}\propto q^2$ in the long wavelength limit  \cite{sinova, baym}. This  contrasts   with the linear spectrum of Bogoliubov phonons in homogeneous BECs \cite{bruderer}.  The so-called  Tkachenko modes \cite{tkachenko}, although heavily damped,  have been detected experimentally in vortex arrays \cite{coddington}.

In this Letter we study a two-component mixture with a large population imbalance, in which a diluted species  plays the role of a mobile impurity while the second, majority  species is prepared in a  vortex lattice state.  In the following we derive the effective impurity-phonon Hamiltonian that describes what we call a  \emph{Tkachenko polaron}. We discuss how the model parameters depend on the density of atoms and vortices as well as the intra- and inter-species interaction strengths. In order to characterize the properties of this new quasiparticle, we calculate the  spectral function $A({\bf k},\epsilon)$ in the weak coupling (``large polaron'') limit by   perturbation theory. In cold atom systems, the spectral function can be measured by momentum-resolved photoemission spectroscopy  \cite{koschorrek}. We find that, in comparison with impurities 
dressed by optical or acoustic phonons, the Tkachenko polaron is distinguished by a  Lorentzian quasiparticle peak with a finite width for arbitrarily small $k$ even at zero temperature. Our main result is that  the decay rate  $\gamma_{\bf k}\propto k^2$ scales linearly with energy,  a characteristic sign of \emph{marginal}  interactions in the sense of the renormalization group (RG) \cite{shankar}. We then calculate the one-loop correction to the effective impurity-phonon interaction and  the mass renormalization. It turns out that the impurity-phonon interaction is marginally relevant, implying that the effective coupling grows as the energy scale decreases and the Tkachenko polaron inevitably becomes  heavy and strongly damped   in the long wavelength limit.

Consider  a  two-component mixture with repulsive contact interactions as described by the Hamiltonian $H=H_A+H_B+H_{int}$ with \bea
H_A&=&\int d^2r\,\left[ \hat\psi_A^\dagger\frac{(-i\hbar\nabla-{\bf A})^2}{2m_A}\hat\psi_A^{\phantom\dagger}+\frac{g_{A}}{2}(\hat\psi_A^{\dagger}\hat\psi_A^{\phantom\dagger})^2\right],\nonumber\\
H_B&=&\int d^2r\,\left[ \hat\psi_B^\dagger\frac{(-i\hbar\nabla)^2}{2m_B}\hat\psi_B^{\phantom\dagger}+\frac{g_{B}}{2}(\hat\psi_B^{\dagger}\hat\psi_B^{\phantom\dagger})^2\right],\label{Hbec}\\
H_{int}&=&g_{AB}\int d^2r\,\hat\psi_A^{\dagger}\hat\psi_A^{\phantom\dagger}\hat\psi_B^{\dagger}\hat\psi_B^{\phantom\dagger}.\nonumber
\eea
Here $\hat \psi_{A,B}(\mathbf r)$  are bosonic field operators for majority ($A$) and impurity ($B$)  atoms with masses $m_A$ and $m_B$, respectively; $g_{A}$, $g_{B}$ are the intra-species interaction strengths and $g_{AB}$ is the inter-species interaction strength. 
The interaction parameters can be tuned by Feshbach resonance. They are related to the three-dimensional $s$-wave scattering lengths $a_j$, for $j=A,B,AB$, by $g_j=2\sqrt{2\pi}\hbar^2 a_j/m_jl_j$, where $m_{AB}=2m_Am_B/(m_A+m_B)$ and   $l_j$ denotes the respective trap lengths in the direction perpendicular to the plane. In addition, $\mathbf A(\mathbf r)$ is an artificial vector potential  \cite{dalibard} corresponding to an effective uniform magnetic field  $\mc B^*$ that  couples selectively to $A$ atoms. The reason for considering artificial gauge fields instead of rotating traps is that we want to induce a vortex lattice in the majority species in the laboratory frame, while the impurity species maintains a nearly free dispersion and interacts only weakly with the local density of $A$ atoms. The creation of vortices in a BEC subject to an artificial magnetic field has been demonstrated experimentally  by Lin et al. \cite{lin}.   In this experiment, a spatially varying  vector potential $\mathbf{A}(\mb r)=-\mc B^*y\hat x$ was engineered via a detuning gradient of Raman beams, but other schemes have  also been proposed  \cite{dalibard}. 
In our model  we are interested in the limit of a large number of vortices $N_V\gg1$ and neglect the effects of a trap potential. The conditions for creating large vortex lattices with artificial gauge fields have been discussed theoretically \cite{dalibard}.

In the polaron problem, we focus on a single $B$ atom interacting with majority atoms with   average two-dimensional density  $n_A=N_A/\mc S$, where $N_A$  is the total number of atoms distributed over an area  $\mc S$. 
Let us  first analyze the state of  the BEC of $A$ atoms. We denote by $\Omega=\mc B^*/m_A$ the cyclotron frequency associated with the  effective magnetic field. In the regime of weak interactions  $g_{A}n_A\ll    \hbar  \Omega$,  i.e. the mean-field quantum Hall regime \cite{ muller,shlyapnikov},  the vortex array  is described by  Gross-Pitaevskii theory. The mean field state  corresponds to condensation  into a single-particle wave function in the lowest Landau level  $\varphi_A(\mathbf r)\propto\prod_\alpha (z-\zeta_\alpha)$. Here $z=(x+iy)/\ell$ is the complex coordinate in the $(x,y)$ plane normalized by the magnetic length $\ell =\sqrt{\hbar/m_A\Omega}$ and $\zeta_\alpha$ are the positions of the vortices arranged in a triangular Abrikosov lattice. The ratio between number of atoms  and number of vortices  is given by the filling factor $\nu=N_A/N_V=\pi n_A \ell^2\gg 1$ \cite{shlyapnikov}. 

The Tkachenko mode  spectrum is calculated in the Bogoliubov approximation by expanding $H_A$ in Eq. (\ref{Hbec}) about the Gross-Pitaevskii solution  to second order in the fluctuations. We write $\hat \psi_A(\mathbf r)\approx \sqrt{n_A}\varphi_A(\mathbf r) +\delta\hat\psi_A(\mathbf r)$ with \be
\delta\hat\psi_A(\mathbf r) = \frac1{\sqrt{ \mc S}} \sum_{\mathbf q\in \textrm{BZ}}\left[u_{\mathbf q}(\mathbf r)a_{\mathbf q}^{\phantom\dagger}-v_{\mathbf q}(\mathbf r)a_{\mathbf q}^{\dagger}\right],\label{deltapsiA}
\ee
where $a_{\mathbf q}$ is the annihilation operator for the Tkachenko mode with wave vector $\mathbf q$ defined in the Brillouin zone of the triangular lattice and  $u_{\mathbf q}(\mathbf r)$, $v_{\mathbf q}(\mathbf r)$ are solutions of the projected Bogoliubov-de Gennes equations \cite{shlyapnikov}. For $q\ll \ell^{-1}$, we can approximate $u_{\mathbf q}(\mathbf r)\approx \varphi_A(\mathbf r) c_{1\mb q}e^{i \mb q\cdot \mb r}$ and $v_{\mathbf q}(\mathbf r)\approx \varphi_A(\mathbf r) c_{2 \mb q}e^{-i \mb q\cdot \mb r}$ where $c_{1 \mb q}\approx \sqrt{(\chi_q/\tilde\omega_q+1)/2}$, $c_{2 \mb q}\approx \sqrt{(\chi_q/\tilde \omega_q-1)/2}$,  $\tilde\omega_q\approx \alpha\sqrt{\eta} (q\ell)^2$, $\chi_q\approx {\alpha[1-(q\ell)^2+(\eta+1)(q\ell)^4/2]}$, with   constants $\alpha \approx 1.1592$ and $\eta\approx 0.8219$. The gapless collective modes (Goldstone bosons of the broken rotational and translational symmetry) have dispersion relation $\hbar \omega_{\mathbf q}\approx \hbar^2q^2/2M$. The ratio between the atomic mass and the Tkachenko mode mass is $m_A/M= 2\alpha\sqrt{\eta} \,n_A g_A/\hbar\Omega\ll 1$. Going  beyond the Bogoliubov approximation, Matveenko and Shlyapnikov \cite{shlyapnikov} showed that  nonlinear corrections  give rise to  Beliaev damping of the Tkachenko modes. However,   the damping rate  is suppressed by a factor of $1/\nu\ll 1$, thus the Tkachenko mode  is a well defined excitation in the mean field  quantum Hall regime.

Let us now turn to the interspecies interaction $H_{int}$ in Eq. (\ref{Hbec}). Expanding to first order in the fluctuation $\delta\hat \psi_A$, we obtain $H_{int}\approx H_{lat}+H_{imp-ph}$. The first term, \be
H_{lat}=\int d^2 r\, V(\mathbf r) \hat\psi^\dagger_B(\mathbf r)\hat\psi^{\phantom\dagger}_B(\mathbf r),\label{potential}
\ee
with $V(\mathbf r)=n_Ag_{AB}|\varphi_A(\mathbf r)|^2$, accounts for  the static lattice potential of the Abrikosov lattice   seen by the impurities. This   is analogous to the periodic potential produced by  laser beams  in optical lattices \cite{bloch}.  But here the  potential stems from the density-density interaction $g_{AB}$, which makes it energetically more  favourable for  $B$ atoms to be located near   vortex cores, where the density of $A$ atoms vanishes. We can compare  the recoil energy $E_R=\hbar^2/2m_B\xi^2$, where $\xi$ is the vortex core size, with the lattice potential depth $V_0=n_Ag_{AB}$.  In the mean field quantum Hall regime, $\xi\sim \ell$ \cite{fetter}, thus $E_R\sim (m_A/m_B)\hbar \Omega$. The  shallow lattice  limit $E_R\gg V_0$ is more natural if $m_A\sim m_B$ and $g_{AB}\sim g_A$. In this work we shall focus on shallow lattices   and derive an effective continuum model.  However,   the deep lattice limit can also be achieved by selecting heavier impurity atoms and by increasing $g_{AB}$.

The combination of the periodic potential  (\ref{potential}) with the kinetic energy in $H_B$ in Eq. (\ref{Hbec}) leads to Bloch bands for the impurity. For weak interactions, we can project  into the lowest  band  and write $\hat\psi_B(\mathbf r)\sim\frac1{\sqrt{\mc S}} \sum_{\mathbf k}\Phi_{\mathbf k}(\mathbf r)b_{\mathbf k}$, where $\Phi_{\mathbf k}$ is a Bloch wave function and $b_{\mathbf k}$ annihilates a $B$ atom with wave vector $\mathbf k$ in the Brillouin zone. In the long wavelength limit, the  lowest band dispersion becomes $\varepsilon_{\mb k}\approx \hbar^2 k^2/2m_B^*$, where $m_B^*$ is the effective impurity mass. Hereafter we set $m_B^*=m$ to lighten the notation. It is interesting to compare $m$ with the Tkachenko mode mass $M$. If  $m\sim m_A$, we expect $m<M$ in the mean field quantum Hall regime. However, the opposite case $m>M$ is also possible in the deep lattice limit as discussed above.

The  term generated by $H_{int}$ to first order in $\delta\hat\psi_A$ is the impurity-phonon interaction. 
Using  the mode expansion in Eq. (\ref{deltapsiA}), we obtain\be
H_{imp-ph}=\frac1{\sqrt{\mathcal S}}\sum_{\mathbf k, \mb q}g_{\mb k,\mb q}b^\dagger_{\mb k+\mb q}b^{\phantom\dagger}_{\mb k}(a^{\phantom\dagger}_{\mb q}+a^{\dagger}_{-\mb q}),\label{Himpph}
\ee
where the impurity-phonon coupling   reads\bea
g_{\mb k,\mb q}&=& \sqrt{n_A}g_{AB}(c_{1\mb q}-c_{2\mb q})\times\nonumber\\
&&\times\int_{s_V} \frac{d^2r}{\pi \ell^2} \,\Phi_{\mb k+\mb q}^*(\mb r) \Phi_{\mb k}(\mb r) |\varphi_A(\mb r)|^2e^{i\mb q\cdot \mb r}.\label{ephcoupling}
\eea
Using the lattice translational symmetry, we have reduced  the integral in Eq. (\ref{ephcoupling}) to the unit cell $s_V$ occupied by a single vortex (with area $\pi\ell^2$). 
In the ``large polaron'' regime  \cite{devreese} we  take the continuum limit  and  expand    Eq. (\ref{ephcoupling}) for   $k,q\ll \ell^{-1}$. The dependence on particle momentum $\mb k$ disappears as the dominant effect in the impurity-phonon interaction  is the slow oscillation of the potential  \cite{herrera}. As a result, the coupling simplifies to \be
g_{\mb k,\mb q}\approx g_{q}=\lambda q, \label{glinearq}
\ee 
where  $\lambda \approx  \eta^{1/4}(\nu/2\pi)^{1/2}g_{AB}.$
The linear momentum dependence in Eq. (\ref{glinearq}) stems from the small-$q$ limit of $c_{1\mb q}-c_{2\mb q}$. For comparison, the same factor yields $g_q\propto \sqrt{q}$ in homogeneous BECs \cite{bruderer,griessner}. 

\begin{figure}
\begin{center}
\includegraphics*[width=.6\columnwidth]{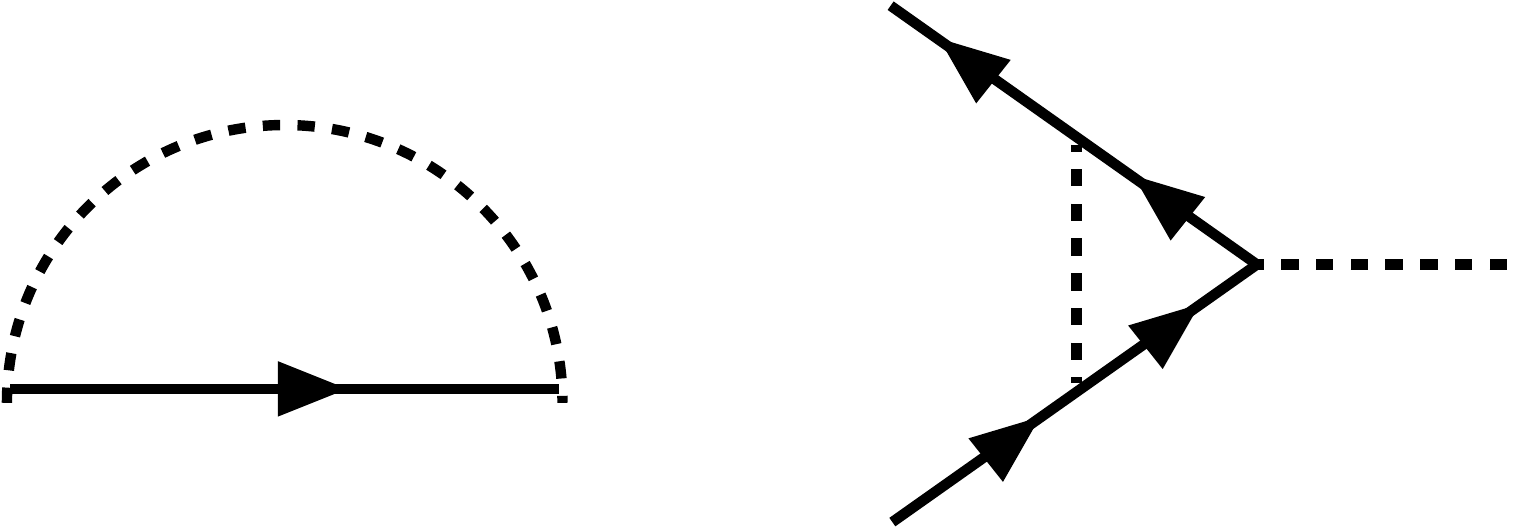}
\end{center}
\caption{Lowest order Feynman diagrams. Left: impurity self-energy. Right: vertex correction to the impurity-phonon interaction. The solid and dashed  lines represent   free impurity and Tkachenko mode (phonon) propagators, respectively.  \label{diagrams}}

\end{figure}

To analyze the effects of the impurity-phonon interaction (\ref{Himpph}), we calculate  the single-particle  spectral function at zero temperature\be 
A(\mb k,\epsilon)=-\frac1\pi\textrm{Im}\left[\frac1{\epsilon-\varepsilon_{\mb k}-\Sigma(\mb k,\epsilon)}\right].\ee
In the weak coupling limit, the lowest order diagram that contributes to the retarded self-energy $\Sigma(\mb k,\epsilon)$ contains one Tkachenko mode in the intermediate state as  illustrated in Fig. \ref{diagrams}. The imaginary part  is given by \be
\textrm{Im}\Sigma(k,\epsilon)=-\frac{ \lambda^2\mu^2}{2\hbar^4}\left(\delta \epsilon+\frac{\mu \hbar^2k^2}{m^2}\right)\theta[\epsilon-\epsilon_{min}(k)],\label{imagpart}
\ee
where  $\delta\epsilon=\epsilon-\hbar^2k^2/2m$, $\mu=mM/(m+M)$, $\theta(x)$ is the Heaviside step function and $\epsilon_{min}(k)$ is the lower threshold of the spectral function imposed by kinematic constraints. The perturbative result gives $\epsilon_{min}(k)=\hbar^2k^2/2(m+M)$, which corresponds to the kinetic energy of the center of mass for two particles with masses $m$ and $M$ and total momentum $\hbar \mb k$. If we go beyond second order and allow for arbitrarily many Tkachenko modes in the intermediate state, we find that $\epsilon_{min}(k)\to 0$ for all $k$. 
The real part of the self-energy reads\be
\textrm{Re}\Sigma(k,\epsilon)=\frac{\lambda^2\mu^2}{\pi\hbar^4}\left[|\delta\epsilon|-\left(\delta \epsilon+\frac{\mu\hbar^2 k^2}{m^2}\right)\ln\left|\frac{\Lambda}{f(k,\epsilon)}\right|\right],\label{realpart}
\ee
where $\Lambda$ is a high-energy cutoff ($\sim E_R$) and $f(k,\epsilon)=(\mu\hbar^2k^2/2m^2)\theta(\delta\epsilon)+[\epsilon-\epsilon_{min}(k)]\theta(-\delta\epsilon)$. We have omitted in Eq. (\ref{realpart}) a constant  term of order $\Lambda$ that amounts to a non-universal     shift in the polaron ground state energy.

\begin{figure}
\begin{center}
\includegraphics*[width=.9\columnwidth]{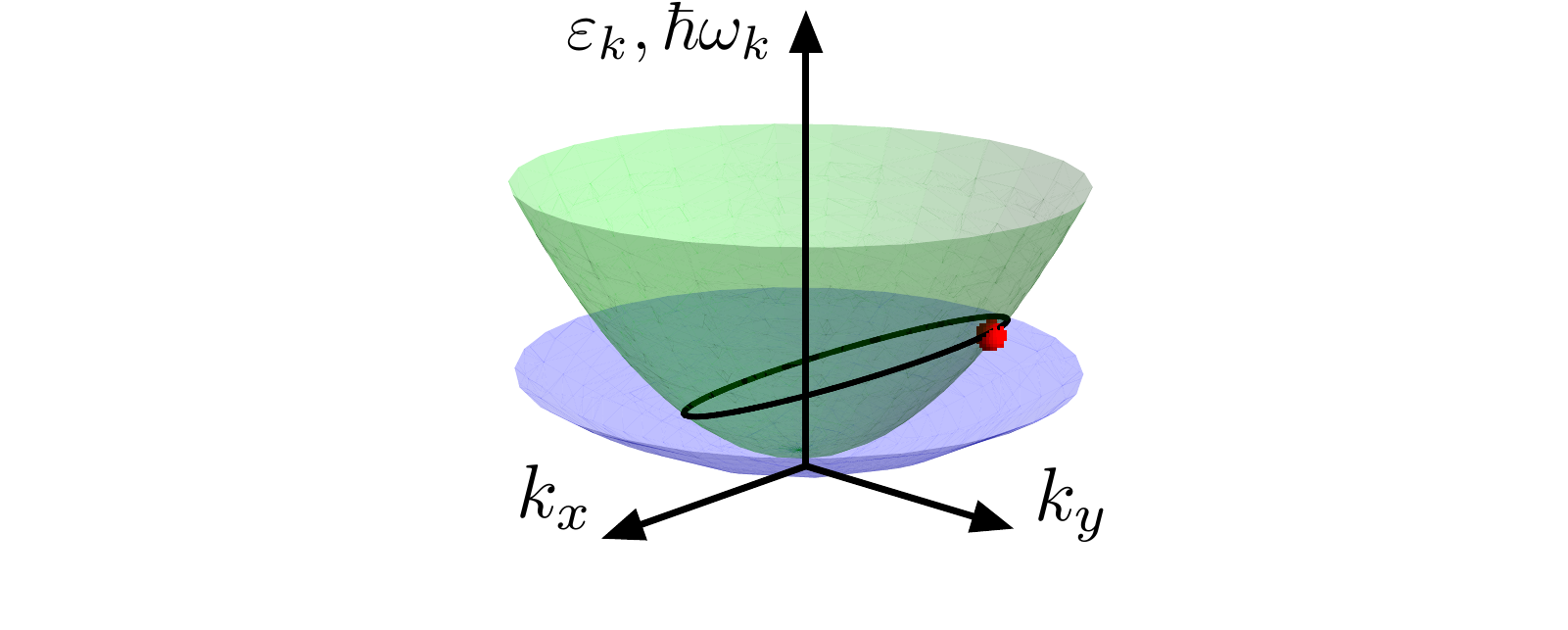}
\end{center}
\caption{(Color online.) Parabolic dispersion of impurity atom  (upper green surface) and Tkachenko mode (lower blue  surface) for mass ratio $m/M<1$. The solid line represents the states into which an  impurity in the state marked by the red dot  can decay   by emitting a Tkachenko mode  respecting  energy and momentum conservation.    \label{paraboloids}}
\end{figure}

Two remarks about Eqs. (\ref{imagpart}) and (\ref{realpart}) are in order. The first remark is that the impurity decay rate $\gamma_{\mb k}\approx -\textrm{Im}\Sigma(k,\varepsilon_{\mb k})\propto k^2$ is nonzero for any $k>0$. This is a direct consequence of the quadratic dispersion of Tkachenko modes, since there is always available phase space for the decay of the single particle respecting energy and momentum conservation (see Fig. \ref{paraboloids}). As a result,   polarons  moving  through the vortex lattice experience a dissipative force.  We stress that the Lorentzian line shape with $\gamma_{\mb k}>0$ at zero temperature  is not observed for small-$k$ particles coupled to acoustic phonons, since in this case the   impurity-phonon continuum lies entirely above the energy of the single-particle state \cite{marsiglio}.  

The second remark is that  the scaling of the decay rate   $\gamma_{\mb k}\sim k^2 \sim \varepsilon_{\mb k}$ signals that the polaron is only marginally coherent. This is consistent with the infrared logarithmic singularity in the real part  Re$\Sigma(k,\varepsilon_{\mb k})\sim \varepsilon_{\mb k}\ln |\Lambda/\varepsilon_{\mb k}|$ for $\varepsilon_{\mb k}\ll \Lambda$. Here \emph{marginal} means that a simple scaling analysis (at tree level \cite{shankar}) predicts that the ratio between the decay rate and the quasiparticle energy does not vary with momentum. 
That the impurity-phonon interaction in Eq. (\ref{glinearq}) is marginal  can be seen by writing down the partition function $Z=\int \mc D\mb X\mc D\phi\,e^{-S[\mb X,\phi]/\hbar}$ with the effective Euclidean action in the scaling limit  \bea
S&=&\int d^2rd\tau\,\left[\frac M2(\partial_\tau\phi)^2+\frac{\hbar^2}{8M}(\nabla^2\phi)^2\right]\nonumber\\
&&+ \int d\tau\, \left[ \frac m2(\partial_\tau \mb X)^2-\frac{\tilde\lambda\hbar^2}{\mu}\nabla^2\phi(\mb X)\right].\label{polaronaction}
\eea
Here $\tilde\lambda=\mu\lambda/\hbar^2$ is the dimensionless coupling constant, $\mb X$ is the position vector of the impurity and $\phi(\mb r)=\sum_{\mb q}(\hbar/2\mc S M\omega_q)^{1/2}(a^{\phantom\dagger}_{\mb q}+a^{\dagger}_{-\mb q})e^{i\mb q\cdot \mb r}$ is a real scalar field defined from Tkachenko  modes. The action in Eq. (\ref{polaronaction}) is invariant under the RG transformation  \cite{shankar} with scale factor $s>1$:  $\mb r^\prime =s^{-1}\mb r$, $\tau^\prime =s^{-z} \tau$, $\phi^\prime=\phi$, $\mb X^\prime =s^{-1}\mb X$, with dynamical exponent $z=2$.

Having established  that the impurity-phonon coupling constant $\tilde \lambda$ is marginal, we  proceed by calculating the quantum correction to scaling. At one-loop level, the renormalization of $\tilde\lambda$ is given by the vertex-correction diagram   in Fig. \ref{diagrams}. Integrating out high-energy modes, we obtain the perturbative RG  equation \be
\frac{d\tilde\lambda}{d\ln(\Lambda_0/\Lambda) }=\frac{\tilde\lambda^3}{\pi}.\label{RG}
\ee
Likewise,   the self-energy diagram yields the  impurity mass renormalization \be
\frac{d  m}{d\ln(\Lambda_0/\Lambda) }=\frac{2\tilde\lambda^2\mu}{\pi}\label{RG2}.
\ee
Interestingly, similar RG equations occur in another two-dimensional system, namely graphene with unscreened Coulomb interactions \cite{guinea}.  But while in graphene the  interaction is marginally  irrelevant,  in our case Eqs. (\ref{RG}) and  (\ref{RG2})  show that $\tilde\lambda$ flows to strong coupling and the polaron mass increases. We interpret this result as a sign that the ``large  polaron'', defined from a small value of the bare coupling constant $\tilde\lambda_0$, becomes  surprisingly heavy and strongly damped at small wave vectors. This effect can be detected experimentally as an anomalous broadening of the spectral function at small $k$ and low enough temperatures  $ T\ll \varepsilon_{\mb k}$.  The precise signature of the marginally relevant interaction is the momentum dependence of  the relative width of the quasiparticle peak.  In the weak coupling regime $e^{-\pi/4\tilde \lambda_0^2}\ll k\ell \ll 1$, the ratio between the decay rate and the energy  \emph{increases} logarithmically with decreasing $k$   (see  Supplemental Material):  
\be
\frac{\gamma_{\mb k}}{\varepsilon_{\mb k}}\approx \frac{\tilde{\lambda}_{0}^{2}}{1+r_0}\left(1+\frac{C\tilde{\lambda}_{0}^{2}}{\pi}\ln\frac{2m_0\Lambda_{0}}{k^2}\right),\ee 
where $C=(5+r_0^2)/(1+r_0)^2$ and $r_0=m_0/M$ is the bare mass ratio.

So far we have neglected the Beliaev damping of Tkachenko modes \cite{shlyapnikov}. Within a perturbative approach with a dressed phonon propagator, it is easy to verify that a finite decay rate for Tkachenko modes leads to further broadening of the polaron spectral function. Remarkably,  the scaling $\Gamma_{\mb q}\propto q^2$ of the Tkachenko mode decay rate  suggests that the nonlinear  (cubic) phonon decay process considered in Ref. \cite{shlyapnikov} is also marginal. However, $\Gamma_{\mb q}/\hbar\omega_{\mb q}\sim 1/\nu$, whereas for the polaron decay rate $\gamma_{\mb k}/\varepsilon_{\mb k}\sim \lambda^2\sim \nu$ according to Eq. (\ref{glinearq}). Therefore, our results are valid in the mean field quantum Hall regime, where $\nu\gg1$.

Finally, we would  like to mention that, although we have focused on the single-particle problem, it should also be interesting to investigate the phase diagram of Tkachenko polarons at finite densities. Furthermore, the strong damping of atoms due to  coupling to Tkachenko modes may be useful 
for  sympathetic cooling if we regard the vortex lattice as a reservoir that can absorb  entropy  and distribute it into long wavelength excitations via multiple decay processes. In fact, the soft parabolic dispersion of Tkachenko modes allows one to circumvent the kinematic constraints that hinder cooling by intraband transitions in the case of acoustic phonons in homogeneous BECs \cite{griessner}.

In conclusion, we have shown that impurities moving in a vortex lattice are strongly damped due to coupling to collective modes with parabolic dispersion. We  derived the effective impurity-phonon model,  which at weak coupling is equivalent to a   quantum field theory with a marginally relevant interaction. As a result, we predict that Tkachenko polarons exhibit an unconventional    decay rate that increases logarithmically with decreasing wave vector. 

We thank  D. Marchand for several discussions and E. Demler and A. Lamacraft for insightful comments. This work is supported by  Fapesp/CEPID (M.A.C., V.S.B.), CNPq-CAPES/INCT, (M.A.C., V.S.B.)  and CNPq grant 309234/2011-5 (R.G.P.).

\onecolumngrid

\appendix
\section{Supplemental Material}
\section{Impurity-phonon interaction for shallow lattice potential}

In the following we set $\hbar=1$. When we substitute the expansion for the field operator $\hat{\psi}_{A}(\mb r)\approx \sqrt{n_A}\varphi_A(\mb r)+\delta \hat{\psi}_{A}(\mb r)$ in $H_{int}$ in Eq. (1) of the main text, the first-order term in the   fluctuation $\delta\hat{\psi}_{A}$ gives the impurity-phonon  interaction 
\begin{eqnarray}
H_{imp-ph} = \sqrt{n_A}g_{AB}\,\int d^{2}r \, \left(\varphi^{\phantom\dagger}_{A}\delta\hat{\psi}_{A}^{\dag}\hat{\psi}_{B}^{\dag}\hat{\psi}^{\phantom\dagger}_{B}+\varphi^*_{A}\delta\hat{\psi}^{\phantom\dagger}_{A}\hat{\psi}_{B}^{\dag}\hat{\psi}^{\phantom\dagger}_{B}\right).
\end{eqnarray}  
For $\delta\hat{\psi}_{A}$, we utilize the mode expansion in Eq. (2) of the main text. The functions $u_{\mb q}(\mb r)$ and $v_{\mb q}(\mb r)$ are obtained following the derivation by Matveenko and Shlyapnikov \cite{shlyapnikov}. We take the continuum limit $q\ell \ll 1$ in Eqs. (22), (23) and (27) of Ref. \cite{shlyapnikov}, which leads to the simplified spatial dependence $u_{\mb q}(\mb r),v_{\mb q}(\mb r)\propto \varphi_A(\mb r)$. In addition, we rescale the momentum $\mb q\to 2\mb q$ in the  definition of the annihilation operators  $a_{\textbf{q}} = \tilde{a}_{2\textbf{q}}$.  We then obtain
\begin{eqnarray}  
H_{imp-ph} = \frac{g_{AB}\, \sqrt{n_A} }{\sqrt{\mc S}} \, \sum_{\textbf{q}} \int \, d^2 r \,   |\varphi_{A} (\textbf{r})|^2  \, e^{i\,\textbf{q}.\textbf{r}} \,  \left(c_{1\textbf{q}}-c_{2\textbf{q}}\right) \,  (a_{\textbf{q}} + a_{-\textbf{q}}^{\dag} ) \, \hat{\psi}_{B}^{\dag}(\textbf{r})\,\hat{\psi}_{B}(\textbf{r}).  \end{eqnarray}  
We expand the field operator $\hat{\psi}_{B}$ in terms of the lowest band Bloch function $\Phi_{\textbf{k}}(\textbf{r})$, which obeys $\Phi_{\textbf{k}}(\textbf{r}+\textbf{R}) = \Phi_{\textbf{k}}(\textbf{r})e^{i\mb k\cdot \mb R}$. Thus 
\begin{eqnarray} 
H_{imp-ph} = \frac{g_{AB}\, \sqrt{n_A} }{\mc S^{3/2}} \, \sum_{\textbf{q},\mb{k},\mb k^\prime}  \left(c_{1\textbf{q}}-c_{2\textbf{q}}\right)  (a_{\textbf{q}} + a_{-\textbf{q}}^{\dag} )  \int d^2 r\,   |\varphi_{A} (\textbf{r})|^2  \, e^{i\textbf{q}\cdot\textbf{r}} \Phi^*_{\mb k^\prime}(\mb r) \Phi_{\mb k}(\mb r). \label{integrals} \end{eqnarray}  

Next, we use the facts  that the local density of majority atoms $|\varphi_A(\mb r)|^2$ is periodic under lattice translations and that     $\Phi_{\mb k}(\mb r)$ is a Bloch function in order to  reduce the integral over the entire system to the integral over a single unit cell $s_V$:\bea
 \int  d^2 r\,   |\varphi_{A} (\textbf{r})|^2  \, e^{i(\mb k+\textbf{q}-\mb k^\prime)\cdot\textbf{r}} \Phi^*_{\mb k^\prime}(\mb r) \Phi_{\mb k}(\mb r)&=& \sum_j\int_{s_V}  d^2 r^\prime\,   |\varphi_{A} (\textbf{r}^\prime+\mb R_j)|^2  \, e^{i\textbf{q}\cdot(\textbf{r}^\prime+\mb R_j)}   \Phi^*_{\mb k^\prime}(\mb r^\prime+\mb R_j)   \Phi_{\mb k}(\mb r^\prime+\mb R_j)\nonumber\\
 &=&\int_{s_V}  d^2 r^\prime\,   |\varphi_{A} (\textbf{r}^\prime)|^2  \, e^{i \textbf{q}\cdot\textbf{r}^\prime} \Phi^*_{\mb k^\prime}(\mb r^\prime) \Phi_{\mb k}(\mb r^\prime)\sum_je^{i(\mb k+\textbf{q}-\mb k^\prime)\cdot\mb R_j} \nonumber\\
 &=&N_V \delta_{\mb k^\prime, \mb k+\mb q}\int_{s_V}  d^2 r^\prime\,   |\varphi_{A} (\textbf{r}^\prime)|^2  e^{i\mb q\cdot \mb r^\prime}  \Phi^*_{\mb k+\mb q}(\mb r^\prime) \Phi_{\mb k}(\mb r^\prime). 
\eea
Thus we can rewrite Eq. (\ref{integrals}) as\be
H_{imp-ph} = \frac{g_{AB}\, \sqrt{n_A} }{\sqrt{\mc S}} \, \sum_{\mb{k},\mb q}\,\mc{I}_{\mb k,\mb q}  \left(c_{1\textbf{q}}-c_{2\textbf{q}}\right)  (a_{\textbf{q}} + a_{-\textbf{q}}^{\dag} )\, b_{\textbf{k}+\textbf{q}}^{\dag} \, b^{\phantom\dag}_{\textbf{k}},
\ee
where (using $N_V/\mc S=1/\pi\ell^2$)\be
\mc I_{\mb k,\mb q}=\int_{s_V}  \frac{d^2 r^\prime}{\pi \ell^2}\,   |\varphi_{A} (\textbf{r}^\prime)|^2  e^{i\mb q\cdot \mb r^\prime}  \Phi^*_{\mb k+\mb q}(\mb r^\prime) \Phi_{\mb k}(\mb r^\prime). 
\ee

In the regime of weak interactions and shallow lattice potential, we approximate the   Bloch functions for small $\mb k$ by plane waves (nearly free impurities), $\Phi_k(\mb r)\approx e^{i\mb k\cdot \mb r}$. In this case, \be
\mc I_{\mb k,\mb q}\approx \int_{s_V}  \frac{d^2 r^\prime}{\pi \ell^2}\,   |\varphi_{A} (\textbf{r}^\prime)|^2  =1, \label{just1}
\ee
where we used the normalization of the ground state wave function $n_A\int d^2r\,|\varphi_A(\mb r)|^2=N_A$. Therefore the dependence on impurity momentum $\mb k$ disappears in the continuum limit. Corrections to Eq. (\ref{just1}) are higher order in momentum or interaction strength. 

Finally, the expansion $c_{1\mb q}-c_{2\mb q} \approx \frac{1}{\sqrt{2}}\,\eta^{1/4}\,(q\,\ell)$ for $q\ell \ll 1$ provides the final expression
\be
H_{imp-ph} = \frac{1 }{\sqrt{\mc S}}\frac{\eta^{1/4}g_{AB} \sqrt{n_A}\ell}{\sqrt2} \, \sum_{\mb{k},\mb q} q  (a_{\textbf{q}} + a_{-\textbf{q}}^{\dag} )\, b_{\textbf{k}+\textbf{q}}^{\dag} \, b^{\phantom\dag}_{\textbf{k}},
\ee
Comparing with Eq. (4) of the main text and substituting $n_A\ell^2=\nu/\pi$, we   identify the impurity-phonon coupling in the ``large polaron'' regime  \begin{eqnarray} \nonumber g_{\textbf{q}} =  \frac{1}{\sqrt{2\pi}} \,\eta^{1/4}g_{AB}\, \nu^{1/2}\,q.
\end{eqnarray}

\section{Effective action for a single impurity}

We start from the Hamiltonian
in second quantization $H=H_{ph}+H_{imp}+H_{imp-ph}$,
\begin{equation}
H_{ph}=\sum_{\mathbf{q}}\frac{q^{2}}{2M}a_{\mathbf{q}}^{\dagger}a_{\mathbf{q}}^{\phantom{\dagger}},
\end{equation}
 
\begin{equation}
H_{imp}=\sum_{\mathbf{k}}\frac{k^{2}}{2m}b_{\mathbf{k}}^{\dagger}b_{\mathbf{k}}^{\phantom{\dagger}},
\end{equation}
\begin{equation}
H_{imp-ph}=\frac{\lambda}{\sqrt{\mathcal{S}}}\sum_{\mathbf{k},\mathbf{q}}qb_{\mathbf{k}+\mathbf{q}}^{\dagger}b_{\mathbf{k}}^{\phantom{\dagger}}(a_{\mathbf{q}}^{\phantom{\dagger}}+a_{-\mathbf{q}}^{\dagger}).
\end{equation}
Switching to first quantization in the Hilbert space of a single impurity,
we can write
\begin{equation}
H_{imp}=\frac{\mathbf{P}^{2}}{2m},
\end{equation}
where $\mathbf{P}$ is the impurity momentum operator. Moreover,
\begin{equation}
H_{imp-ph}=\frac{\lambda}{\sqrt{\mathcal{S}}}\sum_{\mathbf{q}}q(a_{\mathbf{q}}^{\phantom{\dagger}}+a_{-\mathbf{q}}^{\dagger})\sum_{\mathbf{k}}|\mathbf{k}+\mathbf{q}\rangle\langle\mathbf{k}|.\label{eq:1stquantizeHiph}
\end{equation}
The operator acting on the impurity Hilbert space in Eq. (\ref{eq:1stquantizeHiph})
is recognized as $\sum_{\mathbf{k}}|\mathbf{k}+\mathbf{q}\rangle\langle\mathbf{k}|=e^{i\mathbf{q}\cdot\mathbf{X}}$,
where $\mathbf{X}$ is the impurity position operator. Thus we can
write (denoting $\omega_{q}=q^{2}/2M$)
\begin{eqnarray}
H_{imp-ph} & = & \frac{\lambda}{\sqrt{\mathcal{S}}}\sum_{\mathbf{q}}q(a_{\mathbf{q}}^{\phantom{\dagger}}+a_{-\mathbf{q}}^{\dagger})e^{i\mathbf{q}\cdot\mathbf{X}},\nonumber \\
 & = & \frac{\lambda}{\sqrt{\mathcal{S}}}\sum_{\mathbf{q}}\frac{q^{2}}{\sqrt{2M\omega_{q}}}(a_{\mathbf{q}}^{\phantom{\dagger}}+a_{-\mathbf{q}}^{\dagger})e^{i\mathbf{q}\cdot\mathbf{X}}\nonumber \\
 & = & -\lambda\nabla^{2}\phi(\mathbf{X}),
\end{eqnarray}
where we define the dimensionless scalar field from Tkachenko modes
\begin{equation}
\phi(\mathbf{r})=\sum_{\mathbf{q}}\frac{1}{\sqrt{2\mathcal{S}M\omega_{q}}}(a_{\mathbf{q}}^{\phantom{\dagger}}+a_{-\mathbf{q}}^{\dagger})e^{i\mathbf{q}\cdot\mathbf{r}}.
\end{equation}
We also introduce the momentum canonically conjugate to $\phi(\mathbf{r})$
\begin{equation}
\Pi(\mathbf{r})=-i\sum_{\mathbf{q}}\sqrt{\frac{M\omega_{q}}{2\mathcal{S}}}(a_{\mathbf{q}}^{\phantom{\dagger}}-a_{-\mathbf{q}}^{\dagger})e^{i\mathbf{q}\cdot\mathbf{r}},
\end{equation}
so that $[\phi(\mathbf{r}),\Pi(\mathbf{r}^{\prime})]=i\delta(\mathbf{r}-\mathbf{r}^{\prime})$.
The free phonon Hamiltonian can be cast in the field theory form
\begin{equation}
H_{ph}=\frac{1}{2M}\int d^{2}r\left[\Pi^{2}+\frac{1}{4}(\nabla^{2}\phi)^{2}\right]+\mbox{const}.
\end{equation}
Therefore the Hamiltonian for a single impurity coupled to the Tkachenko
field reads
\begin{equation}
H=\frac{\mathbf{P}^{2}}{2M}+\frac{1}{2M}\int d^{2}r\left[\Pi^{2}+\frac{1}{4}(\nabla^{2}\phi)^{2}\right]-\frac{\tilde{\lambda}}{\mu}\nabla^{2}\phi(\mathbf{X}),\label{eq:1stH}
\end{equation}
where $\tilde{\lambda}=\mu\lambda $, with $\mu=mM/(m+M)$, is the dimensionless coupling constant. 

We now switch to a functional integral formalism. A particular configuration
of the system at time $\tau$ is specified by $\mathbf{X}(\tau),\phi(\mathbf{r},\tau)$.
We write $Z=\int\mathcal{D}\mathbf{X}\mathcal{D}\phi\: e^{-S[\mathbf{X},\phi]}$.
The Euclidean action corresponding to Hamiltonian (\ref{eq:1stH})
is
\begin{equation}
S=\int d^{2}rd\tau\left[\frac{M}{2}(\partial_{\tau}\phi)^{2}+\frac{1}{8M}(\nabla^{2}\phi)^{2}\right]+\int d\tau\left[\frac{m}{2}(\partial_{\tau}\mathbf{X})^{2}-\frac{\tilde{\lambda}}{\mu}\nabla^{2}\phi(\mathbf{X})\right].
\end{equation}

\section{Perturbative renormalization group}

In order to derive perturbative RG equations \cite{shankar}, we will need the noninteracting Green's functions for impurity and
phonon operators:
\begin{equation}
G(\mathbf{k},i\omega)=-\int d\tau\: e^{i\omega\tau}\langle T_{\tau}b_{\mathbf{k}}^{\phantom{\dagger}}(\tau)b_{\mathbf{k}}^{\dagger}(0)\rangle_{0}=\frac{1}{i\omega-\varepsilon_{\mathbf{k}}},
\end{equation}
\begin{equation}
D(\mathbf{q},i\nu)=-\int d\tau\: e^{i\nu\tau}\langle T_{\tau}[a_{\mathbf{q}}^{\phantom{\dagger}}(\tau)+a_{-\mathbf{\mathbf{q}}}^{\dagger}(\tau)][a_{\mathbf{q}}^{\dagger}(0)+a_{-\mathbf{q}}^{\phantom{\dagger}}(0)]\rangle_{0}=-\frac{2\omega_{\mathbf{q}}}{\nu^{2}+\omega_{\mathbf{q}}^{2}}.
\end{equation}

\subsection{Interaction vertex}

We define the effective coupling constant $\lambda_{eff}$ from the
three-point function as follows:
\begin{eqnarray}
\Gamma(\mathbf{k},\mathbf{k}^{\prime},\mathbf{q};\omega,\omega^{\prime},\nu) & = & \int d\tau d\tau^{\prime}d\tau^{\prime\prime}\: e^{i\omega\tau}e^{i\omega^{\prime}\tau^{\prime}}e^{i\nu\tau^{\prime\prime}}\langle b_{\mathbf{k}}^{\dagger}(\tau)b_{\mathbf{k}^{\prime}}^{\phantom{\dagger}}(\tau^{\prime})[a_{\mathbf{q}}^{\dagger}(\tau^{\prime\prime})+a_{-\mathbf{q}}^{\phantom{\dagger}}(\tau^{\prime\prime})]e^{-\int d\tau\: H_{imp-ph}(\tau)}\rangle_{0}\nonumber \\
 & = & \frac{1}{\sqrt{\mathcal{S}}}\:\lambda_{eff}q\delta_{\mathbf{k}^{\prime},\mathbf{k}+\mathbf{q}}2\pi\delta(\omega^{\prime}-\omega-\nu)G(\mathbf{k},i\omega)G(\mathbf{k}^{\prime},i\omega^{\prime})D(\mathbf{q},i\nu).\label{eq:lambdaeff}
\end{eqnarray}
To first order in the interaction, we obtain
\begin{eqnarray}
\Gamma^{(1)} & = & -\frac{\lambda}{\sqrt{\mathcal{S}}}\sum_{\mathbf{p}_{1},\mathbf{q}_{1}}q_{1}\int d\tau_{1}d\tau d\tau^{\prime}d\tau^{\prime\prime}\: e^{i\omega\tau}e^{i\omega^{\prime}\tau^{\prime}}e^{i\nu\tau^{\prime\prime}}\langle b_{\mathbf{k}}^{\dagger}(\tau)b_{\mathbf{k}^{\prime}}^{\phantom{\dagger}}(\tau^{\prime})b_{\mathbf{p}_{1}+\mathbf{q}_{1}}^{\dagger}(\tau_{1})b_{\mathbf{p}}^{\phantom{\dagger}}(\tau_{1})\rangle_{0}\nonumber \\
 &  & \times\langle[a_{\mathbf{q}}^{\dagger}(\tau^{\prime\prime})+a_{-\mathbf{\mathbf{q}}}^{\phantom{\dagger}}(\tau^{\prime\prime})][a_{\mathbf{q}_{1}}^{\phantom{\dagger}}(\tau_{1})+a_{-\mathbf{\mathbf{q}}_{1}}^{\dagger}(\tau_{1})]\rangle_{0}\nonumber \\
 & = & \frac{1}{\sqrt{\mathcal{S}}}\:\lambda q\delta_{\mathbf{k}^{\prime},\mathbf{k}+\mathbf{q}}\int d\tau_{1}d\tau d\tau^{\prime}d\tau^{\prime\prime}\: e^{i\omega\tau}e^{i\omega^{\prime}\tau^{\prime}}e^{i\nu\tau^{\prime\prime}}G(\mathbf{k},\tau_{1}-\tau)G(\mathbf{k}^{\prime},\tau^{\prime}-\tau_{1})D(\mathbf{q},\tau_{1}-\tau^{\prime\prime})\nonumber \\
 & = & \frac{1}{\sqrt{\mathcal{S}}}\:\lambda q\delta_{\mathbf{k}^{\prime},\mathbf{k}+\mathbf{q}}2\pi\delta(\omega^{\prime}-\omega-\nu)G(\mathbf{k},i\omega)G(\mathbf{k}^{\prime},i\omega^{\prime})D(\mathbf{q},i\nu).
\end{eqnarray}
Therefore, to first order $\lambda_{eff}=\lambda$. We are interested
in the logarithmic correction to $\lambda_{eff}$ when we integrate
out high-energy modes. At third order (see diagram in Fig. 1 of the
main text), we obtain
\begin{eqnarray}
\Gamma^{(3)} & = & -\frac{\lambda^{3}}{3!\mathcal{S}^{3/2}}\sum_{\mathbf{p}_{1},\mathbf{q}_{1}}q_{1}\sum_{\mathbf{p}_{2},\mathbf{q}_{2}}q_{2}\sum_{\mathbf{p}_{3},\mathbf{q}_{3}}q_{3}\int d\tau_{1}d\tau_{2}d\tau_{3}d\tau d\tau^{\prime}d\tau^{\prime\prime}\times\nonumber \\
 &  & \times\langle b_{\mathbf{k}}^{\dagger}(\tau)b_{\mathbf{k}^{\prime}}^{\phantom{\dagger}}(\tau^{\prime})b_{\mathbf{p}_{1}+\mathbf{q}_{1}}^{\dagger}(\tau_{1})b_{\mathbf{p}_{1}}^{\phantom{\dagger}}(\tau_{1})b_{\mathbf{p}_{2}+\mathbf{q}_{2}}^{\dagger}(\tau_{2})b_{\mathbf{p}_{2}}^{\phantom{\dagger}}(\tau_{2})b_{\mathbf{p}_{3}+\mathbf{q}_{3}}^{\dagger}(\tau_{3})b_{\mathbf{p}_{3}}^{\phantom{\dagger}}(\tau_{3})\rangle_{0}\nonumber \\
 &  & \times\langle[a_{\mathbf{q}}^{\dagger}(\tau^{\prime\prime})+a_{-\mathbf{\mathbf{q}}}^{\phantom{\dagger}}(\tau^{\prime\prime})][a_{\mathbf{q}_{1}}^{\phantom{\dagger}}(\tau_{1})+a_{-\mathbf{\mathbf{q}}_{1}}^{\dagger}(\tau_{1})][a_{\mathbf{q}_{2}}^{\phantom{\dagger}}(\tau_{2})+a_{-\mathbf{\mathbf{q}}_{2}}^{\dagger}(\tau_{2})][a_{\mathbf{q}_{3}}^{\phantom{\dagger}}(\tau_{3})+a_{-\mathbf{\mathbf{q}}_{3}}^{\dagger}(\tau_{3})]\rangle_{0}\nonumber \\
 & = & -\frac{\lambda^{3}}{\mathcal{S}^{3/2}}\: q\delta_{\mathbf{k}^{\prime},\mathbf{k}+\mathbf{q}}2\pi\delta(\omega^{\prime}-\omega-\nu)G(\mathbf{k},i\omega)G(\mathbf{k}^{\prime},i\omega^{\prime})D(\mathbf{q},i\nu)\nonumber \\
 &  & \times\sum_{\mathbf{q}_{1}}q_{1}^{2}\int\frac{d\nu_{1}}{2\pi}\: G(\mathbf{k}+\mathbf{q}_{1},i\omega+i\nu_{1})G(\mathbf{k}+\mathbf{q}+\mathbf{q}_{1},i\omega+i\nu+i\nu_{1})D(\mathbf{q}_{1},i\nu_{1}).
\end{eqnarray}
Comparing with Eq. (\ref{eq:lambdaeff}), we note that
\begin{eqnarray}
\delta\lambda^{(3)} & = & -\lambda^{3}\int\frac{d^{2}q_{1}}{(2\pi)^{2}}\: q_{1}^{2}\int\frac{d\nu_{1}}{2\pi}G(\mathbf{k}+\mathbf{q}_{1},i\omega+i\nu_{1})G(\mathbf{k}+\mathbf{q}+\mathbf{q}_{1},i\omega+i\nu+i\nu_{1})D(\mathbf{q}_{1},i\nu_{1})\nonumber \\
 & = & -\lambda^{3}\int\frac{d^{2}q_{1}}{(2\pi)^{2}}\: q_{1}^{2}\int\frac{d\nu_{1}}{2\pi}\:\frac{1}{(i\omega+i\nu_{1}-\varepsilon_{\mathbf{k}+\mathbf{q}_{1}})(i\omega+i\nu_{1}+i\nu-\varepsilon_{\mathbf{k}+\mathbf{q}+\mathbf{q}_{1}})}\left(\frac{1}{i\nu_{1}-\omega_{\mathbf{q}_{1}}}-\frac{1}{i\nu_{1}+\omega_{\mathbf{q}_{1}}}\right).
\end{eqnarray}
The integration over $\nu_{1}$ (closing the contour in the upper
half of the complex plane) yields
\begin{equation}
\delta\lambda^{(3)}=\lambda^{3}\int\frac{d^{2}q_{1}}{(2\pi)^{2}}\:\frac{q_{1}^{2}}{(i\omega-\omega_{\mathbf{q}_{1}}-\varepsilon_{\mathbf{k}+\mathbf{q}_{1}})(i\omega-\omega_{\mathbf{q}_{1}}+i\nu-\varepsilon_{\mathbf{k}+\mathbf{q}+\mathbf{q}_{1}})}.
\end{equation}
We choose the fast modes to lie in the momentum shell $\mathcal{K}^{\prime}<q_{1}<\mathcal{K}$.
The external momenta and frequencies (for the latter, we perform the
analytic continuation $i\omega\to\omega$, $i\nu\to\nu$) are taken
to be slow, such that $k,q\ll q_{1}$ and $\omega,\nu\ll\omega_{\mathbf{q}_{1}},\varepsilon_{\mathbf{q}_{1}}$.
Thus
\begin{eqnarray}
\delta\lambda^{(3)} & = & \lambda^{3}\int\frac{d^{2}q_{1}}{(2\pi)^{2}}\:\frac{q_{1}^{2}}{(\omega_{\mathbf{q}_{1}}+\varepsilon_{\mathbf{q}_{1}})^{2}}\nonumber \\
 & = & \frac{2\mu^{2}\lambda^{3}}{\pi}\int_{\mathcal{K}^{\prime}}^{\mathcal{K}}\frac{dq_{1}}{q_{1}}\nonumber \\
 & = & \frac{2\mu^{2}\lambda^{3}}{\pi}\ln\frac{\mathcal{K}}{\mathcal{K}^{\prime}}.
\end{eqnarray}
We introduce the cutoff scales with dimensions of energy $\Lambda=\mathcal{K}^{2}/2\mu$,
$\Lambda^{\prime}=(\mathcal{K}^{\prime})^{2}/2\mu$; in this notation,
\[
\delta\lambda^{(3)}=\frac{\mu^{2}\lambda^{3}}{\pi}\ln\frac{\Lambda}{\Lambda^{\prime}}.
\]
Considering an infinitesimal reduction of the cutoff, $\Lambda^{\prime}=\Lambda e^{-dl}$
with $dl\ll1$, we obtain the RG equation for the effective coupling
constant
\begin{equation}
\frac{d\lambda}{dl}=\frac{\mu^{2}\lambda^{3}}{\pi}.
\end{equation}
In terms of the dimensionless coupling constant $\tilde{\lambda}=\mu\lambda$,
we can write
\begin{equation}
\frac{d\tilde{\lambda}}{dl}=\frac{\tilde{\lambda}^{3}}{\pi}.\label{eq:RGflowlambda}
\end{equation}
The $\beta$ function in Eq. (\ref{eq:RGflowlambda}) implies that
$\tilde{\lambda}$ flows to strong coupling. The solution for the renormalized coupling constant is 
\begin{equation}
\tilde{\lambda}^{2}(\Lambda)=\frac{\tilde{\lambda}_{0}^{2}}{1-(2\tilde{\lambda}_{0}^{2}/\pi)\ln(\Lambda_{0}/\Lambda)},\label{eq:solutionlambda}
\end{equation}
where $\lambda_{0}=\lambda(\Lambda_{0})$ is the bare coupling constant.
The perturbative result
breaks down at energy scale $\Lambda\sim\Lambda_{0}e^{-\pi/2\tilde{\lambda}_{0}^{2}}$.
In terms of impurity momentum, the strong coupling regime sets in
below $k\sim\ell^{-1}e^{-\pi/4\tilde{\lambda}_{0}^{2}}$. For small $\tilde\lambda_0$ it may be difficult to observe the full crossover to strong coupling in a finite size vortex lattice since the length scale above which $\tilde\lambda(\Lambda)$ becomes of order 1 is exponentially large.  The line shape of the spectral function at the strong coupling fixed point is an open problem.

\subsection{Anomalous broadening}

The retarded single-particle Green's function for small $k$ and $\epsilon$  can be cast in the standard form\bea
G(\mb k,\epsilon)&=&[\epsilon-\varepsilon_{\mb k}-\textrm{Re}\Sigma(\mb k,\epsilon)-i\textrm{Im}\Sigma(\mb k,\epsilon)]^{-1}\nonumber\\
&\approx&\frac{Z_{\mb k}}{\epsilon-  \varepsilon^*_{\mb k}+i  \gamma^*_{\mb k}}.\label{GwithZ}\eea
At energy scales $\Lambda\gg\Lambda_{0}e^{-\pi/2\lambda_{0}^{2}}$,
we can apply RG improved perturbation theory and replace the bare coupling constants in the lowest order result for the self-energy by the renormalized coupling constants.  We are interested in the logarithmically divergent terms that govern the renormalization  of the quasiparticle weight $Z_{\mb k}$,  of the coupling constant $\lambda$ and of the   effective mass in the dispersion $\varepsilon_{\mb k}^*$. These infrared singularities stem from differentiating the prefactor of  the logarithm in  the real part of the self-energy (Eq. (9) of the main text) with respect to $\epsilon$ and   $k$.  First, we compute the  field renormalization \cite{shankar}\be
Z_{\mb k}^{-1}=1-\left.\left(\frac{\partial\textrm{Re} \Sigma}{\partial \epsilon}\right)\right|_{0}
\approx 1+\frac{\tilde\lambda^2}{\pi}\ln\left(\frac{\Lambda}{\varepsilon_{\mb k}}\right). 
\ee
Therefore the quasiparticle weight $Z_{\mb k}$  decreases logarithmically as $k$ decreases in the weak coupling regime. 

Let us denote by $m_0$ the bare mass (i.e. without logarithmic corrections). The effective mass $m^*$ in $\varepsilon^*_{\mb k}\approx k^2/2m^*$ is related to the self-energy by\be
\frac{m_0}{m^*}=Z_{\mb k}\left[1+m_0\left.\left(\frac{\partial^2\Sigma}{\partial k^2}\right)\right|_0\right].
\ee
To order $\lambda^2$ we obtain\be
\frac{m^*}{m_0}\approx 1-\left.\left(\frac{\partial \Sigma}{\partial \epsilon}\right)\right|_{0}-m_0\left.\left(\frac{\partial^2\Sigma}{\partial k^2}\right)\right|_0
\ee
which gives\be
m^*\approx m_0+\frac{2\tilde\lambda^2\mu}{\pi}\ln\left(\frac{\Lambda}{\varepsilon_{\mb k}}\right).
\ee
Defining the dimensionless mass parameter $\tilde m$ by $m^*=\tilde m \mu$, we obtain\be
\tilde m^*\approx \tilde m_0+\frac{2\tilde\lambda^2}{\pi}\ln\left(\frac{\Lambda}{\varepsilon_{\mb k}}\right).\label{RGmtilde}
\ee 
This is equivalent to the RG equation under an infinitesimal reduction of the cutoff $\Lambda\to\Lambda^\prime e^{-dl}$ \be
\frac{d\tilde m}{dl}=\frac{2\tilde\lambda^2}{\pi}.
\ee

The decay rate in Eq. (\ref{GwithZ}) is given by  \be
\gamma_{\mb k}^*=-Z_{\mb k}\textrm{Im}\Sigma(\mb k, \varepsilon_{\mb k}^*).\label{gammastar}
\ee 
The lowest order result for the imaginary part of the self-energy  (Eq. (8) of the main text) can be written as\be
\textrm{Im}\Sigma(k,\varepsilon_{\mb k})\approx -\frac{\lambda_0^2m_0M^3k^2}{2(m_0+M)^3}.\label{baregamma}
\ee
The logarithmic singularities in Im$\Sigma$ appear at fourth order in perturbation theory. They stem from both vertex corrections (renormalization of $\tilde \lambda$) and self-energy corrections in the internal  impurity line (renormalization of $\tilde m$). We obtain the RG improved decay rate by replacing $\lambda_0$ and $m_0$ in Eq. (\ref{baregamma}) by $\lambda$ and $m$, in addition to including the field renormalization as in Eq. (\ref{gammastar}). The ratio between the renormalized decay rate and the renormalized dispersion becomes\bea
\frac{\gamma^*_{\mb k}}{\varepsilon_{\mb k}^*}&=&\frac{\lambda^2Z_{\mb k} (m^*)^2M^3}{(m^*+M)^3}\nonumber\\
&\approx&\frac{\lambda_0^2m_0^2}{(1+m_0/M)^3}\left[1+\frac{\lambda_0^2m_0^2(5+m_0^2/M^2)}{\pi(1+m_0/M)^4}\ln\frac{\Lambda}{\varepsilon_k}\right],
\eea
where we expanded to first order in the logarithmic correction.
The coefficient of the logarithmic correction is positive for any value of mass ratio $r_0=m_0/M$, thus the relative width  $\gamma_{\mathbf{k}}/\varepsilon_{\mathbf{k}}$ of the quasiparticle  peak increases with decreasing $k$.


\begin{thebibliography}{12}
\bibitem{feynman}L. D. Landau, Phys. Z. Sowjetunion \textbf{3}, 664 (1933); R. P. Feynman, Phys. Rev. \textbf{97}, 660 (1955). 
\bibitem{devreese}J. T. Devreese and A. S. Alexandrov, Rep. Prog. Phys. \textbf{72}, 066501 (2009).
\bibitem{bruderer}M. Bruderer, A. Klein, S. R Clark, and D. Jaksch, Phys. Rev. A \textbf{76}, 011605 (R) (2007).
\bibitem{pupillo}G. Pupillo {\it et al.}, Phys. Rev. Lett.  \textbf{100}, 050402 (2008). 
\bibitem{herrera}F. Herrera, K. W. Madison, R. V. Krems, and M. Berciu, Phys. Rev. Lett. {\bf 110}, 223002 (2013). 
\bibitem{schirotzek}A. Schirotzek, C.-H. Wu, A. Sommer, and M. W. Zwierlein, Phys. Rev. Lett. \textbf{102}, 230402 (2009).
\bibitem{koschorrek}C. Kohstall {\it et al.}, Nature \textbf{485}, 615 (2012); M. Koschorreck {\it et al.}, Nature \textbf{485}, 619 (2012).
\bibitem{marchand}D. J. J. Marchand {\it et al.}, Phys. Rev. Lett. \textbf{105}, 266605 (2010). 
\bibitem{fetter}A. Fetter, Rev. Mod. Phys. \textbf{81}, 647 (2009).
\bibitem{cooper}N. R. Cooper, Adv. Phys. \textbf{57}, 539 (2008). 
\bibitem{sinova}J. Sinova, C. B. Hanna, and A. H. MacDonald, Phys. Rev. Lett. \textbf{89}, 030403 (2002). 
\bibitem{baym}G. Baym, Phys. Rev. Lett. \textbf{91}, 110402 (2003).
\bibitem{tkachenko}V. K. Tkachenko, Sov. Phys. JETP {\bf 22}, 1282 (1966).
\bibitem{coddington}I. Coddington, P. Engels, V. Schweikhard, and E. A. Cornell, Phys. Rev. Lett. \textbf{91}, 100402 (2003). 
\bibitem{shankar}R. Shankar, Rev. Mod. Phys. {\bf 66}, 129 (1994).
\bibitem{dalibard}J. Dalibard, F. Gerbier, G. Juzeliunas, P. \"{O}hberg, Rev. Mod. Phys. {\bf 83}, 1523 (2011).
\bibitem{lin}Y.-J. Lin, R. L. Compton, K. Jim\'enez-Garc\'ia, J. V. Porto, and I. B. Spielman, Nature {\bf 462}, 628 (2009).
\bibitem{muller}E. J. Mueller and T.-L. Ho, Phys. Rev. Lett.  \textbf{88}, 180403 (2002).
\bibitem{shlyapnikov}S. I. Matveenko and G. V. Shlyapnikov, Phys. Rev. A \textbf{83}, 033604 (2011).
\bibitem{bloch}I. Bloch, J. Dalibard,  W. Zwerger, Rev. Mod. Phys. {\bf 80}, 885  (2008).
\bibitem{griessner}A. Griessner, A. J. Daley, S. R. Clark, D. Jaksch, and P. Zoller, New J.  Phys. \textbf{9}, 44 (2007).
\bibitem{marsiglio}Z. Li, C. J. Chandler, and F. Marsiglio, Phys. Rev. B {\bf 83}, 045104 (2011).
\bibitem{guinea}J. Gonz\'alez, F. Guinea, and M. A. H. Vozmediano, Phys. Rev. Lett. {\bf 77}, 3589  (1996).


\end{thebibliography}
\end{document}